\documentclass[12pt,amssymb]{article}
\input epsf.tex
\usepackage{graphicx}
\makeatletter \@addtoreset{equation}{section}

\newcommand{\be}{\begin{equation}}
\newcommand{\ee}{\end{equation}}
\newcommand{\bear}{\begin{eqnarray}}
\newcommand{\eear}{\end{eqnarray}}
\newcommand{\ba}{\begin{array}}
\newcommand{\ea}{\end{array}}

\begin{document}

\begin{titlepage}
\vfill
\begin{flushright}
{\normalsize KIAS-P05037}\\
{\normalsize hep-th/0507257}\\
\end{flushright}

\vfill
\begin{center}
{\Large\bf Reheating the Universe after String Theory Inflation}

\vskip 0.3in

{ Lev Kofman $^{\clubsuit}$\footnote{\tt kofman@cita.utoronto.ca}
and Piljin Yi $^{\spadesuit}$\footnote{{\tt
piljin@kias.re.kr}} }

\vskip 0.15in

{\it  $^{\clubsuit}$Canadian Institute for Theoretical Astrophysics,}\\
{\it 60 St George Str, Toronto, On, M5S 3H8, Canada}\\

 {\it $^{\spadesuit} $School of Physics, Korea Institute for Advanced Study,} \\
{\it 207-43, Cheongryangri-Dong, Dongdaemun-Gu, Seoul 130-722,
Korea}
\\[0.3in]

{\normalsize July  2005}

\end{center}

%\vfill

\begin{abstract}
\normalsize\noindent
In String theory realizations of inflation,
the end point of inflation is often brane-anti brane annihilation.
We consider the processes of reheating of the Standard Model universe
 after brane inflation.
We identify the channels of inflaton energy decay, cascading
from tachyon annihilation through massive closed string loops, KK
modes, and brane displacement moduli  to the lighter standard model
particles. Cosmological data  constrains scenarios by
 putting  stringent limits on the fraction of
 reheating energy  deposited in gravitons and
nonstandard sector massive  relics.
 We estimate the energy deposited into various light
degrees of freedom in the open and closed string sectors, the timing of
reheating, and the reheating temperature. Production of gravitons is
significantly suppressed in warped inflation.  However, we
predict a residual gravitational radiation background at the level
$\Omega_{GW} \sim 10^{-8}$ of the present cosmological energy
density.  We also extend our analysis to multiple throat scenarios.
A viable reheating would be possible in a single throat
or in a certain subclass of multiple throat scenarios of the KKLMMT
type inflation model, but overproduction
of massive KK modes poses a serious problem. The problem is
quite severe if some inner manifold comes with approximate
isometries (angular KK modes) or if there exists a throat of modest
length other than the standard model throat, possibly associated with
some hidden sector (low-lying KK modes).

\end{abstract}

\vfill

\end{titlepage}
\setcounter{footnote}{0}

\baselineskip 18pt \pagebreak
\renewcommand{\thepage}{\arabic{page}}
\tableofcontents
\pagebreak

\section{Generalities: Reheating the Universe}

The transfer of inflaton energy into
radiation energy in the process
of (p)reheating after inflation is a vital part of
any model of early universe inflation.
 According to the inflationary scenario, the universe at early times
expands quasi-exponentially in a vacuum-like state without entropy or
particles. During this stage of inflation, all energy is contained in
a classical slowly moving inflaton field $\phi$.  Eventually  the
inflaton field decays and transfers all of its energy to relativistic
particles, to start the thermal history of the hot Friedmann universe.

The Quantum Field Theory of (p)reheating, i.e. the theory of particle creation
from the inflaton field in an expanding universe,
 describes a process where the quantum effects of particle creation
are not small, but instead produce
a spectacular process where all the particles of the universe are created
from the classical inflaton. The theory of particle creation and
their subsequent
 thermalization  after inflation has a record of theoretical developments
within QFT.
 The four-dimensional QFT
Lagrangian ${\cal L}(\phi, \chi, \psi, A_i, h_{ik}, ...)$ contains the
inflaton part with the potential $V(\phi)$ and other fields which give
subdominant contributions to gravity. In chaotic inflationary models,
soon after the end of inflation the almost homogeneous inflaton field
$\phi(t)$ coherently oscillates with a very large amplitude of the
order of the Planck mass around the minimum of its potential.  Due to
the interactions of other fields with the inflaton in ${\cal L}$, the
inflaton field decays and transfers all of its energy to relativistic
particles.  If the creation of particles is sufficiently slow (for
instance, if the inflaton is coupled only gravitationally to the
matter fields), the decay products simultaneously interact with each
other and come to a state of thermal equilibrium at the reheating
temperature $T_r$. This gradual reheating can be treated with the
perturbative theory of particle creation and thermalization
\cite{Per}.  However, for wide range of couplings
the particle production from a
coherently oscillating inflaton occurs in
the non-perturbative regime of parametric
excitation~\cite{Pre}.  This picture, with variation in its
details, is extended to other inflationary models. For instance, in
hybrid inflation (including $D$-term inflation) inflaton decay
proceeds via a tachyonic instability
of the inhomogeneous modes which accompany symmetry breaking
\cite{Tach}.
One consistent feature of preheating -- non-perturbative copious
particle production immediately after inflation -- is that the process
occurs far away from thermal equilibrium.

 Since hybrid inflation is the closest field theory
prototype of string theory brane inflation, it will be especially instructive to
refer to the theory of tachyonic preheating.
Hybrid inflation involves multiple scalar fields $\Phi$ in the inflaton
sector. It
can be realized in   brane inflation \cite{Brane}, where the inter-brane
 distance is the slow rolling inflaton, while the subsequent  dynamics
of the branes are described by
tachyonic instability in the Higgs direction. Tachyonic instability
 is very efficient so that  the backreaction of
rapidly generated fluctuations does not allow homogeneous background
oscillations to occur because all the energy of the oscillating field is
transferred to the energy of long-wavelength scalar field fluctuations
within a single oscillation.  However, this does not
preclude the subsequent decay of the Higgs and inflaton
inhomogeneities  into
other particles, and thus very fast reheating.
Particles are generated in out-of-equilibrium states with very
 high occupation numbers, well outside of the perturbative regime.

Recent developments in string theory inflation, while in the early stages,
have to address the issue of the end point of inflation, specifically (p)reheating immediately
after inflation. As in the QFT theory of reheating, String Theory reheating
should be compatible with the thermal history of the universe. We can
use this criterion to constrain the
models. Yet, we are especially interested in the
specific String Theory effects of reheating  which may be, in principle,  observationally testable.
>From a theoretical point of view, reheating  after string theory inflation
deals with the theory of particle creation in String Theory, which is an exciting topic by itself.

In this paper we study the transfer of energy into radiation from string theory inflation
based on brane-anti brane annihilation. Brane annihilation is a typical end-point
 of brane inflation.
We assume a "warped" realization of brane inflation, constructed at
the top of the
ground of the KKLT stabilized vacuum \cite{KKLT}.
The models of string theory inflation with warping branch into different possibilities.
We mostly study reheating in the brane-antibrane warped inflation of  \cite{KKLMMT},
but the methods shall be relevant for the  models 
like \cite{eva,fernando}.
Reheating in other versions of string inflation, like
$D3-D7$ inflation of \cite{renata}, potentially
can be described by QFT reheating \cite{spectr},
while the racetrack inflation of \cite{race} relies on field theory entirely.

The warped geometry of string theory inflation has some common features with the
warped  Randall-Sundrum  five dimensional braneworld.
Based on this analogy, reheating in warped string theory inflation was modelled
on the braneworld formalism in the recent paper \cite{barnaby}.

Successful reheating means the complete conversion of inflaton energy into
thermal radiation without any dangerous relics, in order to
provide a thermal history of the universe compatible with
 Big Bang Nucleosynthesis (BBN), baryo/leptogenesis,  and other observations.
Dangerous relics can be massless or massive, and they are each a danger in their own
way. Too many massless relics like radiation of gravitational waves
is excluded by BBN, while too many massive relics overclose the universe.
Therefore we have to monitor undesirable relics in string theory inflationary
scenario. There are significant difference obetween string theory reheating and QFT reheating
in this respect. Indeed,   in string theory we expect excitation
 of all modes interacting gravitationally, gravitational waves, moduli
fields, and
KK modes, and we need their complete decay or extra tuning to go through the needle eye of observational
constraints on potential non-SM particles.

We study reheating after brane-anti-brane annihilation in warped inflation,
more specifically, in the well-known ten dimensional model of Klebanov-Strassler (KS) throat geometry.
We begin with a single throat case to  identify
systematically the channels  of energy transfer
from inflaton to radiation. There are several processes of energy cascading,
from $D-\bar D$ pair annihilation and closed string loop decay to
excitation of KK modes, and
from them to the excitation of open string modes of the SM branes.
 Then we extend our investigation to the multiple throat case,
 associated with different energy scales.
We shall estimate the decay rate of each channel, its energy scales,
the reheating temperature
and the abundance of dangerous relics. Therefore we have to combine the
string theory picture of all relevant  excitations, their QFT  description when it is adequate,
and elements of the theory of early universe thermodynamics. This is a
challenge  for string theorists and cosmologists.

The plan of the paper is the following.

In Section~\ref{sec:KS} we lay-out the background model,
 specifically KS geometry with extra ingredients needed for a
KKLT stabilized string theory solution. We discuss the effect of
warping on the local string modes, which will be major players in the
reheating process.
In  Section~\ref{sec:end} we study qualitatively the cascade of energy from
$D-\bar D$ pairs to radiation.
 Section~\ref{sec:therm} wraps up the thermalization process in the single throat
model.
In Section~\ref{sec:mult} we consider the multiple throat case.

\section{Warped Compactification with Hierarchy}\label{sec:KS}

In finding a realistic model of universe in string theory, one of more
severe constraints come from the so-called moduli problems in cosmology.
If there are light scalar fields around, especially those associated
with moduli fields of compactification, inflation process could easily
read to accumulation of energy in these light modes which then
interferes with either exit from inflation or with low energy
physics at later stage of cosmological evolution.

A simple way out of these moduli problems, which are being investigated
in string theory, involves turning on anti-symmetric tensor fields
along compactified internal dimensions.
In a generic situation with all possible fluxes turned
on and all possible nonperturbative corrections included, it is
believed that the only massless degrees of freedom surviving the
flux compactification would be that of 4-dimensional gravity
or its supersymmetry completion in case of supersymmetric
vacua.

A common feature of these
flux compactification is the warping. In this paper, we will be employing
well-understood case of IIB compactification, where the warping can be
summarized as the ten dimensional metric of the form,
\begin{equation}
G=H^{-1/2}(y)g_{\mu\nu}(x)dx^\mu dx^\nu + H^{1/2}(y){\cal G}_{IJ}dy^I dy^J
\end{equation}
where ${\cal G}$ is a Calabi-Yau metric on 6-dimensional compact
manifold, and the warp factor $H$ depends only on the internal
Calabi-Yau direction.
Note that, as far as the internal manifold goes, this way of writing
the metric is a mere convention. Unless the physical process
concerned depends crucially on underlying supersymmetric structure,
we may as well rewrite the metric as
\begin{equation}
G=H^{-1/2}(y)g_{\mu\nu}dx^\mu dx^\nu + h_{IJ}dy^I dy^J
\end{equation}
since $h\equiv H^{1/2}{\cal G}$ is the relevant physical metric of
the compact direction. This form of the
metric  makes it very clear that the most important
consequence of the warp factor, i.e., a generation of hierarchy
via exponential red-shift is really coming from the $H^{-1/2}$
in front of the first piece.
Precise form of this warp factor is found by solving a second
order equation for $H$ with various source term, such as
contribution from the NS-NS and RR fluxes, D-branes, Orientifold
planes.

\subsection{Klebanov-Strassler Throats}

\begin{figure}
\begin{center}\leavevmode\epsfxsize=0.8\columnwidth \epsfbox{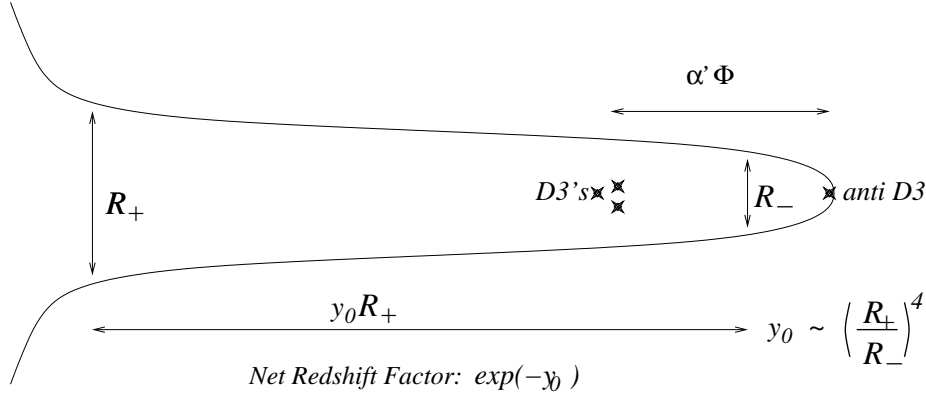}\end{center}
\caption{Radial geometry of a Klebanov Strassler throat.  For most part of this
paper, we consider KKLMMT-like inflation scenario where unstable D-brane system
of D3 branes and anti-D3 branes near the bottom of the throat drives inflation,
possibly with some leftover D3's.}
\label{fig:throat}
\end{figure}

An example of the warped geometry is well-understood
KS solution \cite{Klebanov:2000nc}\cite{KSg}\cite{Frolov:2001xr}.
KS geometry is one of the building block of the KKLT stabilized
solution, which also includes warped instanton branes and $\bar D3$
at the tip of the throat. Using this as a background
(asymptotic of late time cosmological evolution)
inflation can be realized by inclusion of additional elements:
mobile $D3$ brane in the throat attracted to another $\bar D3$
brane around the bottom. This $D3-\bar D3$ interacting pair
provides in inflation in 4d description, end point of inflation
is their annihilation. We also need
 other $D3$ branes around the bottom for SM phenomenology.
Geometry of the model is shown in Figure~\ref{fig:throat}.

The Klebanov-Strassler throat starts from a conifold part of the
Calabi-Yau metric ${\cal G}$. Local form of the metric is
conical,
\begin{equation}
{\cal G}_{IJ}dy^I dy^J=dr^2+ r^2ds^2_{T_{1,1}}
\end{equation}
where the five-dimensional manifold $T^{1,1}$ is a $S^1$ fibred
over a product manifold $S^2\times S^2$. The metric is clearly singular
at origin $r=0$ because the size of $T^{1,1}$ vanishes there.

Actual geometry involves a deformation that blows up one $S^3$,
as a fibration of $S^1$ over one $S^2$, to keep it finite size at
origin while allowing the other $S^2$ collapse to zero size. Thus,
at the bottom $r=0$, the geometry is roughly that of
\begin{equation}
S^3\times R^3
\end{equation}
Corresponding to $S^3$ is a 3-cycle on which RR 3-form flux $F_{(3)}$
is supported, while $R^3$ is part of its dual 3-cycle on which
NS-NS 3-form flux $H_{(3)}$ is supported. Denoting these two
3-cycles as $A$ and $B$, we have the quantization condition
\begin{equation}
\frac{1}{2\pi\alpha'}\int_A F_{(3)}=2\pi M,\qquad
\frac{1}{2\pi\alpha'}\int_B H_{(3)}=2\pi K
\end{equation}
with integers $M$ and $K$.

Note that the $B$ cycle should extend outside of this local
conifold geometry and the quantization condition on  $H_{(3)}$
refers to this entire 3-cycle. However, assuming that most of $H_{(3)}$
flux resides within this conical region, the warp factor has been
solved explicitly. Away from the conifold point $r=0$, we have the following
approximate solution \cite{Giddings:2001yu},
\begin{equation}
H=\frac{1}{r^4}\left(R_+^4+R_-^4\log(r/R_+)^4\right)
\end{equation}
with
\begin{equation}
R_+^4\equiv \frac{27\pi}{4}\alpha'^2g_sMK,\qquad
R_-^4\equiv \frac{3}{8\pi}\frac{27\pi}{4}\alpha'^2g_s^2M^2
\end{equation}
The constant part of $H$ is to be determined by gluing this local
geometry to the rest of the compact Calabi-Yau metric near
$r=R_+$. But for simplicity we have set it to zero.

The two radii, $R_\pm$, are both tunable,
but with string coupling $g_s$ small or with
$K$ sufficiently large, $R_-$ can be considerably smaller
than $R_+$. This is the regime of interest for us since
$R_+> R_-$ typically generate a large hierarchy, be it
between string scale and inflation scale or between
string scale and electroweak scale.

The local geometry in terms of the physical metric is then
\begin{equation}
h\simeq\left(R_+^4+R_-^4\log(r/R_+)^4\right)^{1/2}
\left(\frac{dr^2}{r^2}+ds^2_{T_{1,1}}\right)=\left(R_+^4-4R_-^4y\right)^{1/2}
\left(dy^2+ds^2_{T_{1,1}}\right)
\end{equation}
Up to the change of overall size due to the logarithm,
this metric describes a line parameterized by $y=-\log(r/R_+)$,
times $T^{1,1}$. Together with smooth completion near $r=0$
this is called the Klebanov-Strassler throat. The radius of
$T^{1,1}$ varies from $R_+$ near the top of the throat
($r\sim R_+$) to $R_-$ near the bottom of the throat.
At the bottom of the throat radius of $S^3$ is $R_-$,
while radius of $S^2$ is shrinking to zero.

This approximate metric based on the singular conifold
must be replaced by the one based on deformed conifold
as we reach the size of $T^{1,1}$ to be $R_-$. We will
denote the value of $r$ there as $r_0=R_+e^{-y_0}$, which
takes the value
\begin{equation}
1\gg e^{-y_0}\equiv\frac{r_0}{R_+}=e^{-(R_+^4/4R_-^4)+1}=e^{-(2\pi K/3g_sM)+1}
\end{equation}
For $r<r_0$ the value of $H$ does not change much, and
we then have a redshift factor between the top and the
bottom of the throat
\begin{equation}
1\gg H^{-1/4}(r_0)\simeq \frac{r_0}{R_-}
=\frac{R_+}{R_-}\,{e^{-(R_+^4/4R_-^4)+1}}=\frac{R_+}{R_-}\,{e^{-y_0}}
\end{equation}
This small factor is responsible for generation of hierarchy
between top and bottom of the Klebanov-Strassler throat.

\subsection{Hierarchy and the Decay Cascade of
Local String Modes}\label{sec:cascade}

We identified the cascade of energy transfer from $D-\bar D$ annihilation to SM particles as
shown in the Figure~\ref{fig:cascade}.
First part of the process is creation of closed strings loops.
Then they decay into lighter KK modes.
KK modes interact with SM branes to excite SM particles.
Residual scalar excitations of SM brane decays further into SM fermions.
Each step of this cascade will be described in Section~\ref{sec:end}.
Here we give qualitative description of excitation modes (particles)
in the warp geometry, which will be crucial for quantitative
estimations of Section~\ref{sec:end}.

The hierarchy is generated because the total metric has the
form
\begin{equation}
G=H^{-1/2}g+h
\end{equation}
This structure implies that the conserved energy $E$
in the noncompact spacetime depends on the position of the
quanta in the compact manifold as
\begin{equation}
E^2\sim H^{-1/2}(y)
\end{equation}
This can be seen from the on-shell condition for particle
of mass $m$,
\begin{equation}
H^{1/2} g^{\mu\nu}p_\mu p_\nu + h^{IJ}p_Ip_J=-m^2
\end{equation}
where $p_\mu$ would be the conserved momenta in the spacetime
for a Minkowskii metric $g$.

For degrees of freedom arising from closed strings, the right hand
side would be represented by the oscillator contribution, and we
have
\begin{equation}
-g^{\mu\nu}p_\mu p_\nu = H^{-1/2}(y)\left( h^{IJ}p_Ip_J+
\frac{N}{\alpha'}\right)
\end{equation}
with an integer quantized oscillator number $N$, and the internal
Kaluza-Klein momenta $p_I$. This crude formula is valid only
when we can regard the geometry $h$ and the warp factor $H$ be
both sufficiently slowly varying so that we can regard internal
geometry $h$ nearly flat, and $H$ nearly constant. Furthermore
in this rough scaling argument we also ignore change in string
quantization due to fluxes.

In fact, it is doubtful whether these assumptions are justifiable
for most of closed string modes we are familiar with in old
Calabi-Yau compactification without flux. In the absence of
flux, the Calabi-Yau compactification gives two types of closed
string excitations. One class is the oscillator modes which
we usually ignore for low energy purpose since the mass thereof
are all fundamental string scale $\frac{1}{\sqrt{\alpha'}}$
at least. The other class is
Kaluza-Klein modes which arise from Fourier analysis of 10D
supergravity modes on the compact internal manifold. For large
volume, these latter modes are most relevant. These KK modes
are expressed in terms of eigenmodes of various kind
on internal manifold, which take nontrivial wavefunction
throughout the underlying Calabi-Yau manifold, and can in no way
deemed to be localized in some part of the internal manifold
except for those with extremely large mass.

When the flux compactification involves a warped throat of
the kind we discussed with large redshift factor, however,
a new class of closed string modes emerges. While the conifold
region is a very small piece of the Calabi-Yau manifold,
${\cal G}$,
its associated throat geometry is not necessarily small since the
physical metric is $h=H^{1/2}{\cal G}$ with an exponentially
large $H$. Thanks to this, there are both oscillators mode
and KK modes which are localized deep down the KS throat.
Mass scales of such local modes are
\begin{equation}
m_{KK}\sim H^{-1/4}(r_0)\frac{1}{R_-},
\qquad
m_{oscillator}\sim H^{-1/4}(r_0) \frac{1}{\sqrt{\alpha'}}
\end{equation}
respectively. Provided that $R_-$ somewhat larger than the
string scale $\sqrt{\alpha'}$, it makes sense to consider
such localized and thus redshifted modes.

For these modes at least, the above scaling of on-shell
condition is applicable. While curvature of $h$, gradient of $H$,
and existence of fluxes will change the form of the on-shell
condition, none of them will have as dramatic effect as the
factor $H^{-1/2}$ in front of the right hand side. In terms of
the linear coordinate $y=-\log(r/R_+)$, we have
\begin{equation}
H^{-1/4}\simeq e^{-y}
\end{equation}
up to a prefactor, and this exponential dependence dominates
any kinematics of the closed string modes.

For any massive modes of mass $m^2$ in the unrescaled unit,
located deep down the throat, we have the on-shell condition,
\begin{equation}
-p_\mu p^\mu \sim e^{2y_0-2y} m^2
\end{equation}
with the right hand side increasing exponentially as we
move up the throat, away from conifold toward the bulk
of Calabi-Yau manifold.
This shows roughly how the mass quantization differs
between local modes and the rest. For modes deep down
the KS throat, the mass gap scales either as $\sim e^{-y_0}/R_-$
or as $\sim e^{-y_0}/\sqrt{\alpha'}$, while everywhere else
the mass gap scales as $1/R$, with $R$ being the linear
size of the Calabi-Yau manifold, and $1/\sqrt{\alpha'}$.\footnote{
Here we are assuming that all moduli are fixed by the
flux compactification. For some moduli, notably those
to be fixed by nonperturbative mechanism, the associated
mass scale could be considerably lower. However,
this separation of two scales associated two types
would-be moduli are in principle independent of this
hierarchy we address, and will be taken to be insignificant.}

Therefore, for
energy distributed among local modes to escape the throat,
one must assemble the energy into a few quanta with
exponentially large kinetic energy. The required kinetic
energy must be larger than its rest mass by a factor of
$e^{y_0}\gg 1$. Otherwise, the quanta would be simply
reflected by a Liouville-like wall.
The strength of these Liouville-like walls is dependent on
the mass of the particle: the heavier the local mode, the stiffer
is the wall. Thus, in the presence of this mass-dependent
potential barrier, localized heavy mode will have tendency
to decay near the point $y_c$
\begin{equation}
e^{y_0-y_c}\sim \frac{E}{m}
\end{equation}
to modes with smaller mass scale.

Any heavy localized mode present deep inside the KS throat
would eventually decay to lighter modes within the same throat. As
the energy cascades down to the lighter modes. For $R_-^2$ substantially
larger that $\alpha'$, the lightest modes will be KK modes
localized at the bottom of the throats, and thus in the
intermediate time scale, energy will be deposited in these
modes. While they are light, they also are massive with characteristic
mass scale $e^{-y_0}/R_-$, and thus are also confined within the
throat. The associated potential wall is less stiff, and
the energy can be stored in a somewhat larger volume
at the bottom.

 On a quantitative level, what was said
above can be seen in terms of eigenmodes which obey the
oscillator-like equation \cite{DeWolfe:2002nn,Giddings:2005ff}.
Here instead of writing the full KK mode
equation, we rely on a simple massless scalar field eigenmode
equation
\begin{equation}\label{eigen}
\left[\frac{H(y)}{\sqrt{h}}\;\partial_I \,h^{IJ}
\frac{\sqrt{h}}{H(y)}\,\partial_J\  +H^{1/2}(y) \nabla_\mu \nabla^\mu   \right]\Phi=0
\end{equation}
Factorizing the eigenmode into outer/inner space parts, and replacing
 four-dimensional Lorentz invariant operator $\nabla_\mu \nabla^\mu$ by
its eigenvalue $m^2_{KK}$, we get equation for the KK eigenmode
$\Phi_{m_{KK}}$.
This  $m^2_{KK}$ is KK mass as seen by four-dimensional observer,
$m^2_{KK}=\omega^2-\vec k^2$.
Taking the approximate form
\begin{equation}
H\simeq e^{4y},\qquad
h=R^2\left(dy^2+ds^2_{T_{1,1}}\right)
\end{equation}
where $R$ is a slowly varying function ranging from $R_+$ to
$R_-$ which we take to be constant effectively.
In other words, we can approximate geometry with
that of AdS warped geometry of Randall-Sundrum \cite{Randall:1999ee} type,
times the internal compact manifold $T_{1,1}$ of a definite size.
Singular boundaries  of the Randall-Sundrum geometry are naturally
smoothed out by having the additional internal dimensions. Attachment
to Calabi-Yau manifold and the cigar-like capping of the bottom,
respectively, replace UV and IR branes.

Denoting
the quantized and dimensionless angular momentum on $T_{1,1}$ by $L^2$,
this equation simplifies to
\begin{equation}\label{eigen1}
\left[e^{4y}\partial_ye^{-4y}\partial_y+ m_{KK}^2R^2e^{2y}-L^2\right]\Phi_{m_{KK};L}=0 \ ,
\end{equation}
where $L^2$ is a contribution from angular momentum.
Its spectrum depends on the isometry of $T_{1,1}$.
For instance, contribution from $S^q$
sphere will give $L^2=l(l+q-1)$ where $l$ are integer numbers.

As usual with such warped geometry, the equation can be transformed to
a Bessel equation by taking
\begin{equation}
z\equiv e^{y},\qquad \Phi_{m_{KK};L}=z^2\Psi_{m_{KK};L}
\end{equation}
which gives
\begin{equation}\label{bessel}
\left[\partial_z^2+\frac{1}{z}\,\partial_z+ m_{KK}^2R^2-\frac{4+L^2}{z^2}\right]\Psi_{m_{KK};L}=0
\end{equation}
Thus the eigenmodes to this simplified KK equation are given by linear
combinations of Bessel functions
\begin{equation}\label{besself}
J_{\pm \nu}(m_{KK} R e^y)
\end{equation}
with $\nu^2=4+L^2$  (for $L=0$ we shall take combination of functions
$J_2$ and $Y_2$). This shows that, with the length of the interval in
$z$ coordinate of order $\simeq e^{y_0}$, the mass eigenvalues are
quantized in unit of
\begin{equation}
\Delta m_{KK}\sim \frac{e^{-y_0}}{R}
\end{equation}
as promised. The mass spectrum is roughly
\begin{equation}
m_{KK} \sim n \frac{e^{-y_0}}{R}
\end{equation}
with integers $n$, and for large $n$, the wave function $\Phi_{m_{KK};L}(y)$
is oscillating  near the bottom of the throat, while far away from
the bottom (small $y$) is has the asymptotics
$\Phi_{m_{KK};L}(y) \sim e^{(2\pm\sqrt{4+L^2})y}$.

In the long run, the decay will further proceed until
energy is mostly retained by massless or nearly
massless modes of the string theory in question.
Assuming complete stabilization of Calabi-Yau moduli by flux,
the only such modes are 4 dimensional gravity and possibly light
open string modes associated with stable D-branes which
may exist the bottom of the throat. The question is then
how the final distribution of energy will look like
between the bulk gravitational sector and the localized
open string sector of stable D-branes.

One important point in pursuing this question is that the
only light degrees of freedom is the 4 dimensional gravity,
but this couples to the modes localized at the bottom
of KS throat very weakly. The redshift factor $H^{-1/4}$
reduces the effective scale of energy-momentum by an
exponential factor and pushes down inflation scale and
subsequent reheating scale as well. This is essentially the
physics of Randall-Sundrum scenario, realized in string
theory setting, and can be understood from the fact that
4 dimensional graviton has the wavefunction profile
$H^{-1/2}(y)$ in the internal direction. Any localized mode
at the bottom of KS throat will have very small wavefunction
overlap with 4D graviton and thus cannot generate much gravitational
energy.

On other hand, open strings on D-branes will couple
to KK modes without this exponential suppression but
there still is a volume suppression if they are for
example on D3 branes transverse to the compact Calabi-Yau.
In order to see how much energy is deposited to what species
of particles, we must pay more close attention to interaction
at the bottom of KS throat, which will be discussed in a
later section.

\subsection{Subtleties}

The above equation (\ref{bessel}) is usually
obtained via vast simplification of
actual KK mode equations. In particular, the capping of the
throat near $y=y_0$ is not faithfully reflected, which
leaves determination of precise spectrum difficult. On the
other hand, the robust part of above estimate is that the lowest energy
is of order $\sim e^{-y_0}/R_-$ and also that subsequent gap
between adjacent eigenmodes is also $\sim e^{-y_0}/R_-$.

An important subtlety to bear in mind is how
far one should trust this equal-spaced spectrum. This
linear analysis suggests that a throat of length $R_+y_0$
has an exponentially large number of states of order
$e^{+y_0}$ due to its very small massgap. If we take some
state of mid-range value of the mass $m'$ such that
 $1/R_+\gg m'\gg e^{-y_0}/R_-$, number of states
it can decay to is of order $(m'R_-)e^{y_0}$ and this will
induce a very large width to the eigenmode thus obtained.
For larger enough $m'$, it is therefore reasonable to
expect that this linear analysis is misleading.

This problem also manifests in the shape of the eigenmode.
While it takes a simple innocuous form in the conformal
coordinate $z$, its behavior in physical coordinate
$y$ is much more drastic for high-lying modes. With
$mR_- e^{-y_0}\gg 1$, the Bessel functions oscillate wildly
near $y=y_0$, and its derivative could be larger than the
local string scale. So once we have $m_{KK}$ larger than the
local string scale, it is unclear whether supergravity
approximation can be trusted.

Thankfully, however, analysis of the present work is not be
affected by this ambiguity. We will consider closed string
oscillator modes with perhaps up to 100 oscillators, which
eventually will decay to low lying KK modes. As we will
see later, energy deposited to KK modes will quickly
thermalize among themselves, and since the energy scale
of initial state right after inflation is or order
$\sim 1/4\pi^3 g_s \alpha'^2 < 1/\alpha'^2$, relevant KK modes are those
below string scales.  We never rely on very high scale
KK modes who precise nature would need more careful analysis.
For bulk estimate of reheating processes,
even details of low lying KK modes does not enter other than
their numbers.

One important exception to this is how the
lowest energy eigenvalues depend on the angular momentum
$L^2$. Later we will consider $L^2$ as approximately conserved
quantum number in the full Calabi-Yau compactification,
this precise spectrum at low end could be important for
lifetime estimate of long-lived KK mode carrying such quantum numbers.
We leave it to future study.

\section{Decay of D-Branes at the End of a Brane Inflation}\label{sec:end}

One attractive class stringy inflation models
involve unstable D-brane system \cite{Sen:1998sm}
whose elevated vacuum
energy drives the inflation \cite{Dvali:1998pa}. While introduction of
unstable brane system is an novel element, this difference
does not seem to generate much new flavor in terms
of studying 4 dimensional low energy effective theory
during inflation. As is typical with inflation, maintaining
sufficient number of e-folding and at the same time having
a graceful exit is not a small problem, and these are just
translated to more geometrically constraints on the
underlying string theory.

When it comes to reheating process after the end of inflationary
era, however, behavior of brane
inflation could be very different from ordinary field theory
models. Sometimes the so-called tachyon effective action
is invoked, but one cannot take this tachyon effective
action too literally. While this low energy approach has
been immensely successful, mathematical results one find
of it must be reinterpreted with care. For instance,
the so-called tachyon matter is known to survive the decay
process and takes up all initial energy in the unstable brane
system, and behaves like a perfect fluid of very massive
noninteracting particles. However, the system in question
started with an open string description which should not be
valid by the time D-brane has decayed. This so-called
tachyon matter turns out to be a coarse-grained view
on the underlying physical state, namely a certain
distribution of highly excited closed string states.

In the conventional inflationary models, reheating question
centers on how effective an inflaton decay can excite
other degrees of freedom. Here, the situation is reverse.
The initial reheating process is such that 100\% of the
energy density responsible for the inflationary phase is
converted to heavy degrees of freedom that have nothing
to do what drove the inflation. The right question to ask
here is how this huge amount of energy density is eventually
distributed among different light degrees
of freedom. Since the initial stage of decay produces a lot
of heavy closed strings, there is an inherent danger of closed
strings dominating the process. Assuming some kind of
brane worlds scenario for standard model sector, this
would be a disaster for brane inflation to be viable, or
looked backward a very efficient and simple tool for
eliminating many stringy inflation scenarios.

In this section, we will describe a very efficient and viable
reheating process for the case of single throat scenario.
In later section we will discuss under what circumstances
multi-throat scenario can offer a viable reheating process.

\begin{figure}
\leavevmode\epsfxsize=\columnwidth \epsfbox{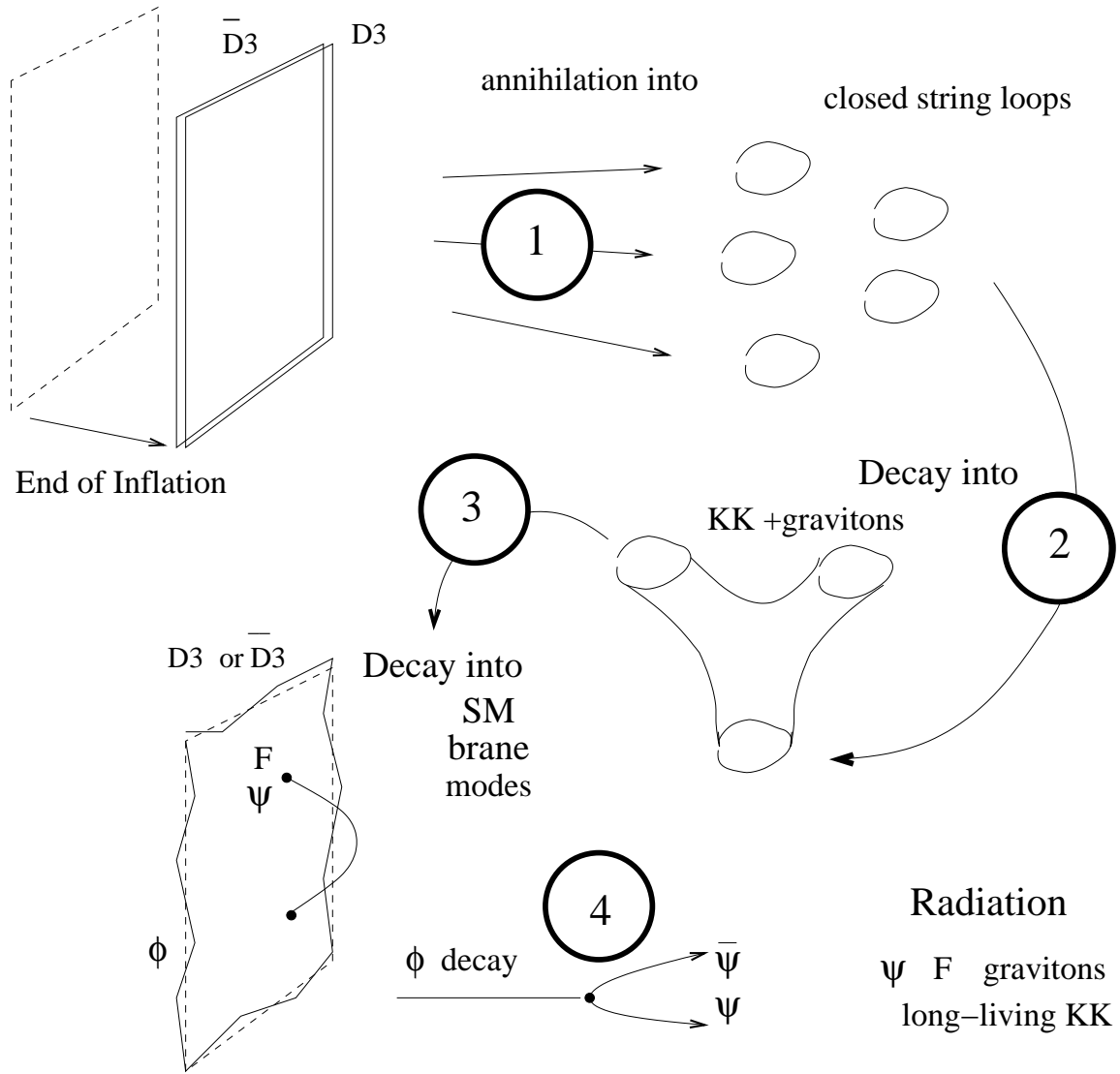}
\caption{Identifying the channels of D-brane decay}
\label{fig:cascade}
\end{figure}

\subsection{$D-\bar D$ Annihilation and Closed Strings Production}\label{sec:annihilation}

The end point of inflation is $D3-\bar D3$ pair annihilation, the step 1 at
 the Figure~\ref{fig:cascade}.
Complete description of non-BPS system in string theory is a complicated
problem \cite{Sen}.
The hallmark of the D-brane decay, as opposed to decay of
inflaton in field theoretical models of inflation,
is the fast and complete conversion of unstable D-brane energy into
massive closed string modes \cite{Yi:1999hd,Bergman:2000xf,Sen:2003xs}.
This is what we find from tachyon dynamics in low energy
\cite{Gibbons:2000hf,Sen:2000kd,Sen:2002an,Gibbons:2002tv}
but also supported by stringy computation \cite{Chen:2002fp,LLM,Gutperle:2004be}
using decaying boundary state \cite{Sen:2002nu,Sen:2002in,Mukhopadhyay:2002en,Rey:2003xs}.

The pair of isolated $D3-\bar D3$ brains annihilates into
excitations of close strings  loops with the average energy $E$,
 in the bosonic string theory  \cite{LLM}  $ E$ is
\begin{equation}\label{LLM}
\frac{E}{V_3} \simeq  \sum_N \int d^{(d-3)}k_{\perp} \, D(N) \, e^{-2\pi \omega_{N,k}} \ ,
\end{equation}
where $\omega_{n,k}=\sqrt{\vec k_{\perp}^2+4N}$,
 $V_3$ is the three dimensional volume of the branes, $\vec k_{\perp}$ is
the momentum of the closed strings transverse to the branes, $D(n)$ is the
number of closed strings oscillator states, for large $N$
\begin{equation}\label{D}
D(N) \simeq N^{-q} e^{4\pi\sqrt{N}} \ ,
\end{equation}
in bosonic theory $q=27/4$, $d=25$; in superstring theory $d=9$ and $q$ is not known.

It is interesting to compare this result with the inflaton decay in the QFT tachyonic
preheating, which is characterized by the high occupation numbers.
In brane-anti brane annihilation, the product  states are characterized by two
numbers, oscillator state $n$ and momentum $k$. analog of the occupation number would be
$ e^{-2\pi \omega_{N,k}}$ which is less than unity.
However, QFT has no analog of the  oscillator number. If we sum over $n$ for a given $\vec k$,
in principle we can get the number exceeding unity.

With a view toward reheating from brane inflation, we will not need much of
the details of the decay. Let us summarize the main
characteristics of closed strings from decay of unstable D-branes:

Initially massive oscillator modes are produced. The probability of
a particular closed string mode with energy $ \omega_{N,k}$ to be produced scales
as $e^{-2\pi \omega_{N,k}}$,
but this exponential suppression is exactly cancelled
by the Hagedorn growth of density of state $D(N)$ at larger energy. The
upshot is that for each oscillator level $N$, roughly the same amount
of energy is deposited.

Since boundary state is formulated at $g_s\rightarrow 0$ limit, one
must introduce cut-off to emulate backreaction of the
boundary state to production of closed strings. For unstable D0, this
natural cut-off is $m_s/g_s$ with the local
string scale $m_s$, and with this cut-off
the energy in the produced heavy closed string account for all
energy in the unstable D-branes.
The highest  oscillator stare which is expected to be excited
is estimated from $\omega_{max} \sim 1/g_s$.
For the string coupling $g_s \sim 0.1$ we get $N_{max} \sim 100$.

The probability distribution $e^{-2\pi \omega_{N,k}}$ also implies a narrow velocity
dispersion. The usual string on-shell condition $\omega \sim \sqrt{p^2+m_s^2 \, N}$,
implies that the transverse velocity of typical closed string mode
is at most of order $\sim \sqrt{m_s}/\sqrt{m}\sim \sqrt{g_s}$. Thus,
kinetic energy of the produced closed strings tends to be smaller than
its rest mass.
The string loops are non-relativistic, only very slowly can move
around their birth place in the volume $V_{d-3}$.

If the unstable D-brane system decays well inside the KS throat
with a redshift factor $H^{-1/4}(r_0)$ and a large radius $R_-$,
all of above should remain true qualitatively. The main difference
is in the string scale $m_s$. Since we are discussing energy
in terms of 4-dimensional metric $g$, the string scale $m_s$ that
appear above should be related to the fundamental string tension
by
\begin{equation}
m_s^2=\frac{H^{-1/4}(r_0)}{2\pi\alpha'}\sim e^{-2y_0}\frac{1}{\alpha'}
\end{equation}
Thus, the decay of D-brane at the bottom of a KS throat produces
strings of mass $m$ and energy $E$ of order $e^{-y_0}/(g_s\sqrt{\alpha'})$
at most.

{}From the four dimensional perspective, energy density of the closed loops
is the same as the energy density of the  $D3-\bar D3$ pair.
We have to take into account redshift of the brane tension in the warped geometry
by factor $e^{-4 y_0}$
\begin{equation}\label{density}
\epsilon = 2T_3 e^{-4 y_0}=\frac{e^{-4 y_0}}{ 4\pi^3 g_s \alpha'^2} \ ,
\end{equation}
which is the scale of the warped inflation.

The timing of $D3-\bar D3$ annihilation we estimate as
\begin{equation}\label{time1}
\Delta t_1 \sim e^{y_0}\sqrt{\alpha'} \ .
\end{equation}

Strictly speaking, the string computation was done
for decay process involving 1-point function of closed strings,
and for this reason can be relied on only for unstable D0 or
unstable D-branes wrapped on a small torus.
The same computation for more generic cases is inconclusive
because the 1-point emission is not the dominant decay
channel. In fact, energy
output from such 1-point emission can be computed and can be
shown to be well short of the expected energy output.
Mathematically this happens because the produced string mode
cannot carry momentum longitudinal to the D-brane. However,
this is an artifact of 1-point emission processes.

For higher dimensional cases, the dominant process should
involve simultaneous production of two or more closed strings,
which has more phase space volume along the longitudinal
direction. Although such a multi-point amplitude
would be suppressed by powers of string couplings, the gain from
phase space volume can easily overcome this. For instance,
consider an unstable D1 which is infinitely extended.
Two point decay would be suppressed by one more power
of $g_s$, but the available phase volume grows linearly
with the energy of the created pair of particles. Introducing
a natural cut-off $1/g_s$ in string scale, the phase volume will
be of order $1/g_s$. This cancels the coupling suppression from
the string diagram easily; we expect this process to
dominate the decay process with the quantitatively same
characteristics of produced closed strings.

Consideration of low energy approach using tachyon effective
action has been developed independent of the string theory
computation and tells us pretty much the same story. The
so-called tachyon matter \cite{Sen:2002nu,Sen:2002in} and string fluid
\cite{Gibbons:2000hf,Sen:2000kd,Gibbons:2002tv,Kwon:2003qn}, which emerges
from this low energy approach, have been studied in depth and compared
to closed string side \cite{Sen:2003xs,Yee:2004ec}, for unstable
D-branes of arbitrary dimensions. The result shows that all energy
is converted into heavy closed string excitations, possibly together
with long fundamental strings.

\subsection{Closed Strings Decay to Local KK Modes}\label{sec:closed}

Now that we identified the initial state right after
the end of brane inflation, we must consider the
subsequent decay of massive closed string thus produced.
This corresponds to the step 2 at the Figure~\ref{fig:cascade}.

The local closed string modes will then decay to whichever
are lighter degrees of freedom around. There are two types of
light modes at the bottom of throats. One are KK modes, which
would be lighter that oscillator modes, as long as $R_-$ is
not too near the string scale. Another are light open strings modes
associated with extra D-branes. One easy way to achieve this
is to have some extra D3-branes surviving at
the end of annihilation. While decay amplitude to a pair of open string
modes would carry one less factors of $g_s^{1/2}$, it is also
suppressed by volume effect $\sim (\alpha'/R_-^2)^{3/2}$.
Even with $R_-/\sqrt{\alpha'}$ slightly larger 1, the latter
effect will easily compensate for the former, and will favor
decay to KK modes first. In this work, we assume that $R_-$
is  large enough to justify field theory and gravity
analysis employed here. Even if it is necessary to extrapolate to
smaller value $R_-$, our guess is that basic qualitative estimates
we offer in this paper would remain valid. After all, what
really matters at the end of day is that energy settles in
some open string sector that would contain the standard model,
and non-SM dangerous relics.

In the flat background massive closed string with oscillation
 number $n$ decays into
two closed strings with oscillation numbers $N-N'$ and $N'$, $N' < N$,
with the coupling strength $g_s$. Subsequently, $N-N'$ and $N'$
 states further cascade
into states with the smaller oscillator numbers.
The final states will be KK modes, fraction of which is in
massless gravitational radiation in the bulk.
Gravi-tensor projection of the bulk gravitons to
four dimensions give describe usual gravitational waves
in four dimensions.
Therefore,
 in the inflationary scenarios which have
$D-\bar D$ annihilation at the end, we shall address
the problem of overproduction of gravitational radiation.
One of the possibility to overcome the problem is to arrange
annihilation not between a $D-\bar D$ pair, but a system of
$\bar D$ and several $D$ branes, to dilute to amount of gravitons.

However, decay of closed strings in the warped geometry is significantly
different from that in the flat background.
Although the energy of the unstable D-brane system
is all deposited to the highly excited strings, none
of these modes can overcome the potential barrier
toward the top of KS throat, since there is $e^{y_0}$
factor difference between strings scales between the
top and the bottom.

The only exception to this
would be the part of energy radiated into massless
modes. In the current setup the only massless
graviton is the 4-dimensional variety, whose wavefunction
comes with the same exponential suppression at the bottom
of the throat. Another way to say this is that the
effective tension of the unstable D-brane is
redshifted by a power of $e^{-y_0}$ relative to the Planck
scale which weakens its coupling to the 4 dimensional gravity.
It also can be seen in the following way.
The wave function of massless four dimensional graviton
(zero KK mode), after rescaling with respect to the warp factor,
 is homogeneous relatively to inner dimensions.
Therefore decay of the closed strings into gravitons is
suppressed by the four dimensional Planck mass,
while decay into massive KK modes is suppressed
by the redshifted string scale.
Therefore, we have very specific effect of the warped geometry of
 the exponentially suppressed
production of gravitational waves from
$D-\bar D$ annihilation.
To the first approximation, no energy is deposited to
the 4 dimensional graviton directly.
More specifically, the energy fraction deposit in gravitational
waves radiation is $e^{-2y_0}$ of the total energy of radiation.
This figure depends on the energy scale of inflation.
Recall that amount of energy density in gravitational waves at the moment of BBN
shall be no more than several percent
of the total radiation  energy density.

Apart from the 4
dimensional gravitons, the coupling to which is universally
small deep inside the throat, the next light degrees of
freedom are local KK modes. To see this clearly and also
for a later purpose of computing branching ratio, let us
expand the 10 dimensional Einstein action according to
the compactification. Starting with 10 dimensional
action,
\begin{equation}\label{ten}
\frac{1}{2g_s^2\kappa_{10}^2}\int dX^{9+1} \sqrt{-G} R_G
\end{equation}
with $\kappa_{10}^2=(2\pi)^7\alpha'^4/2$, take as before
$G=H^{-1/2}g+h$ but allow small fluctuations in $g$ and
$h$. The former gives the 4-dimensional gravitons while
the latter is prototype for KK modes from the compact
manifold. Expanding the 10 dimensional action and keeping
terms quadratic to the small fluctuations, the first term
\begin{equation}
\frac{1}{16\pi G_N}
\int dx^{3+1} \sqrt{-g}R_g
\end{equation}
is the Einstein action in 4 dimensions with the
effective Newton's constant $G_N$ such that
\begin{equation}
\frac{1}{16\pi G_N}=\frac{\tilde V_6}{2g_s^2\kappa_{10}^2},
\qquad \tilde V_6\equiv
\int dy^5\sqrt{h} H^{-1/2}
\end{equation}
Note that the KS throat region contribution to $\sim \tilde V_6$
is of order $R_+^6$, so we have a bound $V_6>R_+^6$. Large
$R_\pm$ will necessarily make the volume of the compact manifold
somewhat larger than fundamental string scale, and weaken
gravity further, regardless of the exponential redshift factor.

In our convention, the gravitational redshift due to the warp factor
is manifest on the matter effective action. If we expand
$h=h_{KS}+\delta h$, the remaining terms has the following general
structure,
\begin{eqnarray}
\frac{1}{2g_s^2\kappa_{10}^2}\int dx^{3+1}
\sqrt{-g}\left(\int dy^{6} \sqrt{-h}\, H^{-1/2}\,
\delta h \nabla^2_{g}\,\delta h +\int dy^6 \sqrt{h}
\,H^{-1}\,\delta h \nabla^2_{h_{KS}}\,\delta h\right)
\end{eqnarray}
which are the kinetic term and the mass term, respectively, of
internal KK modes. Further expanding $\delta h$ in terms
of eigenmodes localized at the bottom of the throat,
\begin{equation}
\delta h=\sum \phi^{(n)}_{KK}(x)\delta h_{(n)}(y)+\cdots
\end{equation}
where $\delta h_{(n)} $ denote local KK modes,
all supported at the bottom $r\simeq r_0$.

As a matter of convenience we will normalize $\delta h_{(n)}$
to produce canonical kinetic term for $\phi_{KK}^{(n)}$, so that
these second class of terms produce,
\begin{equation}
\frac{1}{2}\int dx^{3+1}\sqrt{-g} \sum_n \left( (\nabla \phi^{(n)}_{KK})^2
-(m_{KK}^{(n)})^2(\phi_{KK}^{(n)})^2\right)
\end{equation}
The KK masses $m_{KK}^{(n)}$ of the localized modes are of order
\begin{equation}
m_{KK}^{(n)}\sim e^{-y_{(n)}}\frac{1}{R_-}
\end{equation}
where we introduced $y{(n)}$ whereby $\delta h_{(n)}$ occupy
region at the bottom of the throat $y>y_{(n)}$.

As we saw above, the initial product of D-brane decay is numerous
heavy closed strings of mass $\sim e^{-y_0}/g_s\sqrt{\alpha'}$
or smaller. With not too small $R_-$, local KK modes are much lighter
and more numerous than these local oscillator modes, and will couple  to
the closed string modes in usual 3-point diagram to these oscillator
modes with the coupling being essentially $g_s$. The KK modes thus
produced can have energy or mass up to $\sim e^{-y_0}/g_s\sqrt{\alpha'}$,
which implies that the energy is distributed in KK quanta and excites
the throat up to $y_{0}-y \sim \log(R_-/g_s\sqrt{\alpha'})$.

In contrast, the
4 dimensional graviton, which is the only light mode that are not
localized at the bottom of the throat, has much higher mass scale
of its own,
\begin{equation}
M_P\simeq \frac{1}{g_s} \frac{\sqrt{V_6}}{\kappa_{10}}
\sim  \frac{1}{g_s}\frac{R_+^3}{(\alpha')^{3/2}}\frac{\varepsilon}{\sqrt{\alpha'}}
\sim  \left( \frac{27\pi M K}{4 g_s^{1/3}}  \right)^{3/4} \, \frac{\varepsilon}{\sqrt{\alpha'}}
\end{equation}
which is exponentially larger by a factor of $e^{y_0}$
than that of other scattering processes among local string
modes and local KK modes. Here the numerical constant $\varepsilon$
which came from the definition of $\kappa_{10}$ is
\begin{equation}
\varepsilon\equiv\left(\frac{2}{(2\pi)^7}\right)^{1/2}
\end{equation}
Up to this stage, the primary decay
channel of closed string produced from the D-brane decay will be
into local KK modes.

Finally we estimate the timing of closed strings cascading into light KK modes.
Rate of decay of the individual close string is about
$\Gamma \sim g_s^2 e^{-y_0}/\sqrt{\alpha'}$.
The longest process is the closed string with the oscillator number $N_{max}$
goes through  $N_{max}$ decays, so the upper bound on the net decay time is
$\Delta t_2 \sim N_{max}/ \Gamma$. Since $N_{max} \sim 1/g_s^2$, we estimate
\begin{equation}\label{t2}
\Delta t_2 \sim g_s^{-4} e^{y_0} \, \sqrt{\alpha'} \ .
\end{equation}
his timing is longer than the brane annihilation $\Delta t_1$.
KK mods are ultrarelativistic. Indeed, closed strings have energy
$m_{closed} \sim \frac{e^{-y_0}}{g_s \sqrt{\alpha'}}$, while
KK modes have the mass
$m_{KK} \sim \frac{\sqrt{\alpha'}}{R_-} \, \frac{e^{-y_0}} {\sqrt{\alpha'}}$, so that
$\frac{m_{closed}}{m_{KK}} \sim \frac{R_-}{\sqrt{\alpha'}} \, \frac{1}{g_s}$.

Next we proceed with the decay of KK mode
into open strings modes, which can be associated with the SM particles.

\subsection{KK Modes Decay to Open String Modes}\label{sec:KKdecay}

Let us note that upon exciting KK modes, the corresponding metric
perturbation along the compact direction has the size
\begin{equation}
\delta h(y) \sim \frac{g_s \kappa_{10} H^{1/4}(r_0)}{R_-^3}\,
\phi_{KK}^{(n)}
\end{equation}
for each KK eigenmode with the canonically normalized 4 dimensional
massive fields $\phi_{KK}^{(n)}$. This is what couples to open string
degrees of freedom directly as we will see shortly.
With this, let us consider how KK modes couple and decay to
open string modes on a D-brane transverse to the Calabi-Yau
direction.

One may be alarmed to see the exponentially
large factor $H^{1/4}$ in this expression, but all this does is
to rescale the dimensionful parameters in terms of the local
redshifted scale, since
\begin{equation}
\frac{\kappa_{10} H^{1/4}(r_0)}{R_-^3}
\sim \frac{\alpha'^2 H^{1/4}(r_0)}{R_-^3}
\simeq\frac{(e^{-y_0}/R_-)^3}{(e^{-y_0}/\sqrt{\alpha'})^4 }
\end{equation}
The mass scale of relevant $\phi_{KK}$'s are anywhere
between $e^{-y_0}/R_-$ and $e^{-y_0}/\sqrt{\alpha'}$, so
as long as $R_-$ is not too small in string unit, the
fluctuation is small and treatable as a perturbation.

With this in mind, let us consider a probe  D3 brane located at the
bottom of the tip. The Born-Infeld action can be written as,
\begin{eqnarray}
-\frac{1}{8\pi^3 g_s  \alpha'^2}\int dx^{3+1} H^{-1}(r)
\sqrt{-{\rm Det}\left(g_{\mu\nu}+ 2\pi \alpha' F_{\mu\nu}+
H^{1/2}(r)h_{IJ}\partial_\mu Y^I
\partial_\nu Y^J\right)}
\end{eqnarray}
where $r=r(Y)$ and  $Y^I(x)$ represent transverse fluctuation of D3 brane
along the compact directions.
The primary interaction between
$\delta h$ and D-brane appears from the leading expansion,
\begin{equation}
\frac{1}{8\pi^3 g_s  \alpha'^2} \int dx^{3+1}
\sqrt{-g}\,\left(H^{-1/2}(Y)\left(\frac12
h_{IJ}\partial_\mu Y^I \partial_\nu Y^J g^{\mu\nu}\right) - H^{-1}(Y)
\right)
\end{equation}
Here we have chosen the longitudinal coordinate system such that
\begin{equation}
g_{\mu\nu}=\eta_{\mu\nu}+\delta g_{\mu\nu}
\end{equation}
where $\delta g$ represents the 4-dimensional graviton. For the
moment we set $\delta g =0$ so that $g=\eta$.

In addition, there is another potential term from the
minimal coupling to the background R-R 4-form potential
$C_4$. For this let us recall that KS geometry
comes with 3 types of fluxes; NS-NS 3-form flux $H_3=dB_2$,
R-R 3-form flux $F_3$, and  5-form flux which is related
to the previous two by
\begin{equation}
F_5=dC_4+B_2\wedge F_3
\end{equation}
and is constrained to be self-dual. A minimal $C_4$ for this
self-duality constraint to hold can be determined from $B_2$ and
$F_3$. Relying on the explicit solution of Ref.[KS], we find
\begin{equation}
C_4 =c H^{-1}\,dx^0\wedge dx^1\wedge dx^2\wedge dx^3
\end{equation}
D3 branes have the usual minimal coupling to $C_4$. The
proportionality constant $c$ is such that this contribution
from R-R coupling cancels the tension term from the Born-Infeld
piece exactly for a probe D3 brane and double it for probe anti-D3
brane.\footnote{This is the same cancellation/doubling that is usually
employed  in deriving a slow role potential
for a D3-anti-D3 pair.} Thus, the potential term $H^{-1}$ would
be either cancelled or doubled in actual dynamics.

Expanded up to quadratic order in $Y$, and
canonically normalizing $Y$ fields to $\hat Y$,
we have the following general form of action
\begin{equation}
\frac12\int dx^{3+1}\, \sqrt{-g}\,\left((\partial \hat Y)^2- \mu^2 \hat Y^2
+\frac{(e^{-y_0}/R_-)^3}{(e^{-y_0}/\sqrt{\alpha'})^4 }
\,\phi_{KK}(\partial \hat Y)^2\right)
\end{equation}
where we ignored a multiplicative constant of order one
in front of the last, cubic interaction term. The mass term
is $\mu_{\bar D3} \sim e^{-y_0}/R_-$ for anti-D3 brane, but
for D3 brane, $\mu_{D3}=0$ within the present framework.

With a single throat containing both inflation and standard
model, there should be a further hierarchy generating mechanism.
One such possibility is to have supersymmetry unbroken to much lower
energy scale. For this, we must then consider leftover D3's
rather than anti-D3's, for otherwise we would end up breaking
supersymmetry at the inflation scale. The story of SM  brane phenomenology
is out of the scope of this paper, we just assume
small but non-zero $\mu$ which is larger than TeV scale.

 In fact, it is not even necessary to assume
that the transverse scalar fields are the main decay channel for
the KK modes into SM brane world. {\it The above three-point coupling
is  generic enough to work on any low lying degrees of
freedom on the SM brane world.} For instance, worldvolume fermions
would also have such a 3-point coupling, and can absorb energy from
KK modes directly. Since we are only interested in rough estimate of
reheating process, we will work with the scalar $Y$ and its specific
form of coupling while keeping in mind that the result applies to
all standard model sector. The interaction between $\phi_{KK}$ and $Y$
has the general form
\begin{equation}\label{inter}
\frac{1}{\Lambda} \,\phi_{KK}(\partial \hat Y)^2
\end{equation}
with
\begin{equation}\label{inter1}
\frac{1}{\Lambda}= e^{y_0} \, \frac{\alpha'^2}{R_-^3}=
\left( \frac{71 g_s^2 M^2}{32}\right)^{3/4} \, e^{y_0} \, \sqrt{\alpha'} \ .
\end{equation}
This type of  interaction is typical for interaction
of the radion in the brane-worlds scenarios
\cite{Phenom}.
The rate of decay of KK modes into $Y$ is
\begin{equation}\label{gamma}
\Gamma_{KK}=\frac{m_{KK}^3}{32\pi \Lambda^2} \, \sqrt{1-\frac{4 \mu^2}{m_{KK}^2}} \ .
\end{equation}
For anti-D3 brane we have a problem of decaying into $Y$ particles, since
 $\mu_{\bar D3}$ and $m_{KK}$ are comparable.

On the other hand the mass $\mu_{\bar D3}$ can be arranged to be smaller than
 $m_{KK}$, and therefore D3 branes as the SM brane is preferable.
We proceed assuming
Thus the timing of
decay of KK modes is $\Delta t_3=1/\Gamma_{KK}$
\begin{equation}\label{t3}
\Delta t_3=32 \pi \, \left(\frac{R_-}{\sqrt{\alpha'}}\right)^9 \, e^{y_0} \,\sqrt{\alpha'}=
32 \pi \left(\frac{71 g_s^2 M^2}{32}\right)^{9/4} \,  e^{y_0} \,   \sqrt{\alpha'}
\end{equation}
If we adopt $R_-$ to be larger than $\sqrt{\alpha'}$,
then so far this is the longest process in the chain of energy transfer
from inflaton field.
As we encounter many times throughout the paper, once again
our conclusion depends on specific range of parameters.
Our choice here is based on
 justification of
computation in low curvature KS geometry. As we mentioned above,
KK modes can also decay directly into SM fermions, with the rate of
decay similar to (\ref{gamma}).

Before proceeding with the decay open string modes into SM particle,
we shall discuss separately a special case of
KK modes, which can be long living and most dangerous for the whole scenario.

\section{Thermalization  and Dangerous KK Relics}\label{sec:therm}

In the previous section we consider decay of massive (but light) KK modes into
excitations of the SM brane, and it was found that in the model with
radius of KS throat around its tip $R_-$ significantly  larger than $\sqrt{\alpha'}$
(as expected in the supergravity description of KS geometry),
decay time of KK modes (\ref{t3})  is significant.

KK modes are self-interacting. Therefore
let us check if  KK modes are thermalized or not
 before they
decay into open string modes. In this section we address thermalization of KK modes.
Then we consider
residual decay of open string modes into SM particles, and thermalization of SM particles.
Most importantly, we identify the problem of long-living KK modes, which
is a  serious problem for the string inflation scenarios.

\subsection{Thermalization of KK modes}\label{subsec:therm}

To check whether or nor  KK modes are thermalized
  we have to compare
the time of relaxation of KK modes towards their thermal equilibrium, $\tau$,
and time of their decay $\Delta t_3$.

Relaxation time towards thermal equilibrium is
\begin{equation}\label{relax}
\tau \sim \frac{1}{ n \sigma v} \ ,
\end{equation}
where $n$ is the 3d number density of KK modes,
$\sigma$ is cross-section of their rescattering and $v$ is their typical velocity.
We have to estimate each factor in (\ref{relax}).

In principle each factor in (\ref{relax}) depends on the
 expansion of the universe.
We however argue it can be ignored.
Indeed, let us estimate the value of the Hubble parameter $H$ immediately after
brane-anti brane annihilation. From the Einstein equation
\begin{equation}\label{hubble}
3H^2 \simeq \frac{1}{M_P^2} \,  \frac{e^{-4 y_0}}{ 4\pi^3 g_s \alpha'^2} \ ,
\end{equation}
where we use equation (\ref{density}) for the energy density of the universe
at the end of inflation. From here
\begin{equation}\label{hubble1}
H \sim \frac{e^{-2y_0}}{\sqrt{12 \pi^3 g_s}} \,  \frac{1/\sqrt{\alpha'}}{M_P} \, \frac{1}{\sqrt{\alpha'}} \ .
\end{equation}
The inverse Hubble parameter, which is a typical time of expansion,  is suppressed by the small factor $e^{-2y_0}$,
 while all time intervals
of interaction including $\tau$, as we will see  below, are suppressed by factor $e^{-y_0}$ only.
Therefore here expansion  can be ignored.

Number density can be estimated by ratio of
the total energy density $\epsilon$ and energy per KK mode.
  $\epsilon$ can be taken to be the energy density
 after the brane annihilation (\ref{density}), and energy per KK mode
to be comparable with the  energy
of closed string loops $e^{-y_0}/\sqrt{\alpha'}$, from which they originated.
Thus
\begin{equation}\label{n}
n \sim \frac{e^{-3y_0}}{4 \pi^3 g_s \alpha'^{3/2}} \ .
\end{equation}
KK modes have masses $m_{KK} \sim \frac{e^{-y_0}}{R_-}$ while
they are created from closed strings of the mass $\frac{e^{-y_0}}{\sqrt{\alpha'}}$,
so that KK modes are     relativistic    $v=1$.

 More elaborated is estimation of
$\sigma$. Interaction of KK modes can be derived in the following
way. Again, take 10d action (\ref{ten}) with the ansatz
$G=H^{-1/2}g+h$ as before in Section \ref{sec:closed}. The four
dimensional action, extended up to non-linear terms with respect
to $h$, contains the term $\frac{1}{M_P^2} \int d^{3+1}x \sqrt{-g}
(h \nabla h)^2$. After decomposition $h=h_{KS}+\delta h$ it
contains, in principle, three legs and four legs interactions,
which are comparable. We take four-leg interaction $(\delta h
\nabla \delta h)^2$. Cross section of this interaction $\sigma$ is
the ratio of  squares of coupling
    and   energy, the coupling
is $\frac{g_s^2 e^{2y_0} k^2}{\alpha'^2}$, thus coupling is rather strong $\sim g_s^2$.
This is because KK modes are very energetic, $k \sim \frac{e^{-y_0}}{\sqrt{\alpha'}}$,
 so that momentum factors compensate
the suppression by the local string mass.
The estimate of relaxation time is
\begin{equation}\label{relax1}
\tau \sim \frac{4 \pi^3}{g_s} e^y{_0} \sqrt{\alpha'} \ .
\end{equation}
Relaxation time of KK modes rescattering is much shorter than the time of decay
of unstable KK modes $\Delta t_3$.
Therefore we
can treat all KK modes as particles in thermal equilibrium.
We encounter rather unusual situation in cosmology when\\
{\it  non-SM particles  are set in the thermal equilibrium first, even before
SM particles are produced!}

Therefore the issue of the reheat temperature is split into two issues:
what is the reheat temperature of KK particles, and what is the reheat temperature
of SM particles.

To calculate reheat temperature of KK particles, $T_{KK}$, we have to
convert the energy density KK particles into thermal energy of KK  plasma.
 Energy density of KK particles
\begin{equation}\label{tempk}
\epsilon \simeq c \, \frac{  M_P^2}{ t^2} \ ,
\end{equation}
where numerical coefficient $c$ depends on equation of state, for radiation equation of state
of KK particles $c=\frac{3}{4}$.
The end of inflation fixes the initial  value of $t=t_0$ in (\ref{tempk}).
We have to compare  (\ref{tempk}) energy density (\ref{tempk})
at the moment $t=t_0+\tau$ with thermal energy of KK particles.
Suppose KK particles are thermalized being relativistic.
Then their energy is
\begin{equation}\label{tempk1}
\epsilon=\frac{\pi^2}{30} g_{KK} T_{KK}^4 \ ,
\end{equation}
where $g_{KK}$ is a number of degrees of freedom of KK particles.

A subtle  point is that $t_0 \gg \tau$.
Indeed,
the end of inflation $t_0$ is defined by equalizing (\ref{tempk}) with the energy-density of
original $D3-\bar D3$ pair (\ref{density})
\begin{equation}\label{tempkb}
\epsilon = c \, \frac{  M_P^2}{ t_0^2} =\frac{e^{-4 y_0}}{ 4\pi^3 g_s \alpha'^2} \ ,
\end{equation}
or similarly from $t_0 \sim 1/H$ with $H$ from (\ref{hubble1}).
We find
\begin{equation}\label{t0}
t_0 \sim \sqrt{4\pi^3 c g_s} \, e^{2y_0} \, \frac{M_P}{1/\sqrt{\alpha'}} \, {\sqrt{\alpha'}}
\end{equation}
thus
$t_0 \gg \tau$.

Therefore, comparing $ c \, \frac{  M_P^2}{ t_0^2}$ with (\ref{tempk1}),
we have
\begin{equation}\label{tempk2}
T_{KK} \sim  \frac{e^{-y_0}}{g_s^{1/4}g^{1/4}_{KK}} \, \frac{1}{\sqrt{\alpha'}} \ .
\end{equation}
Thus, reheat temperature of KK modes up to a factor $g_{KK}^{-1/4}$ is comparable with the inflaton
energy scale.

It is instructive to compare $T_{KK}$ with the mass of KK particles
$m_{KK} \sim \frac{e^{-y_0}}{R_-}$. Unless $g_{KK}$ is very large, reheat temperature
of KK particles is larger than their mass, so they are thermalized as relativistic particles.
A new interesting element here is $g_{KK}$ factor. As we seen in
Section 3.2,  KK modes have energies up to $e^{-y_0}/g_s \,
\sqrt{\alpha'}$, the number of degrees of KK modes $g_{KK}$ of the
mass $m_{KK}$ is then proportional to $R_-/g_s \, \sqrt{\alpha'}$.
However, formula (\ref{tempk1}) is valid only for relativistic KK
modes, i.e. for modes with masses less than $T_KK$. For
non-relativistic KK modes we have to use formula
$\epsilon_{KK}=m_{MM} \ , n_{KK}$ with $n_{KK}=\left(\frac{m_{KK}
T}{2\pi}\right)^{3/2} \, e^{-m_{KK}/T}$.

\subsection{Disappearance of Open String Relic}

 Among the product of decaying KK modes into open string sector of SM brane,
there was the scalar $Y$, describing displacement of SM brane. Specific
field content of such scalars are dependent on how SM is realized, but
nevertheless we must worry about such particles as undesirable moduli
fields in cosmological sense. We need the mechanism to give rid off $Y$,
even if such a field exists on SM brane world.
Fortunately, the gauge theory associated with the brane
contains fermions $\psi$ in the adjoint representation.
Decay of $\phi_{KK}$ can go also into fermions, we are interested here in
the troublemakers $Y$.
The scalars $Y$ interact with the fermions
with the three-linear coupling $\sqrt{g_s} \bar \psi Y \psi$.
Therefore $Y$ shall completely decay into lighter fermions, the rate of decay is
\begin{equation}\label{gamma3}
\Gamma (Y \to \bar \psi \psi)=\frac{g_s  \mu_{D3}}{8 \pi} \ .
\end{equation}

Remember that the mass $ \mu_{D3}$ here is introduced on the phenomenological ground beyond the
brane inflation scenario. For  $ \mu_{D3}$ exceeding Tev scale
cosmological constraints are satisfied with a large margin.
Thus decay of the scalars $Y$ by itself is not essential process of the reheating, as far as
SM phenomenology from string theory
is successful.

\subsection{Reheating of SM Particles}

Let us calculate the reheat temperature of the SM universe
in the scenario.
The longest process in the chain of energy transfer  is the
decay of KK modes into open string modes given by time
$\Delta t_3$ (\ref{t3}).
Now we can compute the reheat temperature of SM sector
after the warped brane inflation.

For this we have to convert the energy density stored in the thermal
radiation of KK modes
\begin{equation}\label{temp}
\epsilon_{KK}=\frac{3 M_P^2}{4 (t_0+\Delta t_3)^2}
\end{equation}
into thermal energy of SM plasma
\begin{equation}\label{temp1}
\epsilon_{SM}=\frac{\pi^2}{30} g_* T^4 \ .
\end{equation}

Again, as in the Section 4.1, we shall compare $t_0$ from (\ref{t0}) with the
decay time of KK modes $\Delta t_3$ from   (\ref{t3}).
The answer is model-
dependent.
In principle, $R_-$ can be as small as string scale, see e.g discussion in \cite{CMP}.
In this case $\Delta t_3$ can be smaller than  $t_0$ and
estimation of SM particles temperature is similar to that
of KK modes of Section 4.1, just with replacement of $ g_{KK}$.
However, in the toy model of KS geometry in supergravity limit
with $R_- \geq 10 \sqrt{\alpha}$,
for inflation at energy scale $10^{14} Gev$ with $e^{-y_0} \sim 10^{-4}$,
 we have  $\Delta t_3$ greater than $t_0$ by a large margin.

Energy transfer completed when the Hubble parameter $H(t) \sim 1/t$ drops
below the rate $\Gamma_{KK}$, i.e. when $t$  in (\ref{temp1})
is equal to  $\Delta t_3$.
In (\ref{temp1}) $g_*$ is the number of SM degrees of freedom, $g_* \sim 100$.
We have
\begin{equation}\label{temp2}
T_r \simeq 0.1 \sqrt{M_P \Gamma_{KK}} \sim
 0.01 \, e^{-y_0/2} \, \left( \frac{\sqrt{\alpha'}}{R_-}\right)^{9/2} \, \sqrt{ \frac{1/\sqrt{\alpha'}}{M_P}} \, M_P \ .
\end{equation}
This estimation is not sensitive to the equation of state for KK
modes, if they would have the matter equation of state, it changes
only numerical coefficients in (\ref{temp}). Reheat temperature
(\ref{temp2}) is suppressed by two small factors, the redshift
factor $e^{-y_0/2}$, and by the ratio of scales
$\frac{\sqrt{\alpha'}}{R_-}$. Reheat temperature of SM sector is
 lower the reheat temperature (\ref{tempk2})  of KK
particles

Suppose the energy scale of brane inflation at $10^{14} GeV$, $e^{-y_0}\sim 10^{-4}$, and geometry is such that
$R_- \sim 10 \sqrt{\alpha'}$. We get $T_r \sim 10^7 Gev$.
Now
suppose the brane inflation is the low energy Tev scale inflation with $e^{-y_0} \sim 10^{-15}$,
and  $R_- \sim 10 \sqrt{\alpha'}$. Then $T_r \sim 100 Gev$, which is cosmologically acceptable.
If we choose however, $R_- \leq 100 \sqrt{\alpha'}$, reheat temperature drops
below Mev scale, the temperature of BBN, which is unacceptable.
Reheat temperature in the scenario is generally lower than that in
QFT inflation.
Therefore we conclude that \\
{\it the space of warped brane inflation is  constrained from
the lowest value of reheat temperature}.

\subsection{Problem of Long-Living KK modes}

Above we silently assumed that KK modes all decay into open string modes through
three-legs interaction (\ref{inter}) which make the decay complete.
This  however is not true for a special class of KK modes for which three-legs
interactions like (\ref{inter}) are forbidden.
Now we are coming to the critical problem of the whole scenario we identified so far,
the problem of long-living  KK modes.
They are massive so that after universe expansion dilutes energy density
of radiation they become dominant. Unless extreme fine-tuning, these particles
will have unacceptably large energy density $\Omega_{KK}$.

In the particular model of warped inflation with KS geometry,
there is exact isometries $S^3$ and $S^2$
  of six dimensional interior manifold $X^6$ at the bottom of the throat.
Suppose more generally we have $S^q$ isometry inside $X^6$, and $\theta_{A}$ are
angular coordinates of $S^q$.
KK modes will be described by the harmonic expansion of rank two tensor with respect to
the eigenmodes $\delta h^{(L)}(\theta_A)$
\begin{equation}\label{sphere}
\nabla_A \nabla^{A}  h^{(L)}  =L^2 \,  h^{(L)} \ ,
\end{equation}
 (we drop $\delta $ from $\delta h$ for transparency),
with the Laplace operator on $S^q$ in the right hand side of the equation.
The eigenmodes  $\delta h^{(L)}$ are characterized by the set of conserved quantum numbers,
associated with the conservation of angular momentum and its projections.
For example, for the eigenmode of spin $s$ and  $S^3$ with isometry $SO(4) \simeq SU(2) \times SU(2)$,
we have \cite{Deger}\\
$L^2=(l+1)^2-(s+1)$.

The amplitudes of interactions of  KK modes include three-legs decay interactions with the factor
$\int d\Omega_q  \,  h^{(L)}$ which   vanishes. The only non-vanished amplitudes are
the annihilation and inverse annihilation
amplitudes $\int d\Omega_q   h^{(L)}  h^{(L')} $
which conserve the angular momentum.

In the KS geometry, around the bottom of the throat we have
isometry of $S^3$ sphere with the radius $R_-$. As far as we place
SM brane at $S^3$, the whole $S^3$ isometry is distorted by the
presence of $D3$ brane, and some quantum number of the original
$SO(3)$ isometry are not precisely conserved. However, residual
$SO(2)$ isometry is intact. Another two directions have isometry
of $S^2$  but its radius shrinks.
 This would blow up corresponding KK mass, so that corresponding KK modes will stay away
from the bottom.

Rescattering  of KK modes with angular momentum brings them into the thermal equilibrium
with the rest of KK sector. However, in an expanding universe
 the number density of  KK modes with angular momentum freezes out.
Indeed,
 massive KK particles sooner or later becomes
non-relativistic.
Number density of non-relativistic particles
$n_{KK}=\left(\frac{m_{KK} T}{2\pi}\right)^{3/2} \, e^{-m_{KK}/T}$
is decreasing exponentially as temperature $T$ diluted with expansion,
rate of annihilation and inverse annihilation is exponentially decreasing,
and when it becomes comparable with the expansion rate of the universe,
the abundance of long-living KK modes freezes out.

The problem of KK modes with angular momentum  in KS geometry has similarity with the
problem of heavy KK modes in supergravity noted in \cite{KS}.

Suppose the angular KK modes are stable. Then
the figure of the primary interest would be  the present day abundance $\Omega_{KK}$ of the stable KK modes.
To estimate  $\Omega_{KK}$, we have to use the standard theory of freeze-out species in expanding
universe, see e.g. \cite{KT}.
In principle. particles may freeze out being relativistic or non-relativistic,
depending on the details of the model.
Simplest estimation of  $\Omega_{KK}$ is for relativistic freeze out.
In this case we have
\begin{equation}\label{omega1}
\Omega_{KK} \simeq 0.16 \, \frac{g_{st}}{g_{KK}} \, \frac{m_{KK}}{eV}
\end{equation}
where $g_{st}$ and $g_{KK}$ are the numbers of degrees of freedom of
freezed out stable KK modes and
all KK modes correspondingly.
The value of $\Omega_{KK}$ is of order  $10^{22}$ for the $10^{14}$ GeV scale inflationary throat.

However, the isometries one find in KS throat is only approximate in
the context of Calabi-Yau compactification,
and therefore there is no absolutely stable KK angular modes..
 While a KS throat that
extends infinitely to UV region do have exact $SU(2)\times SU(2)$ isometry,
the actual internal manifold involves cutting-off the UV region
by a finite Calabi-Yau manifold. This has the effect of distorting
the small $r$ part of the KS throat and destroy the isometries.
This then propagates toward IR end of the throats in such a way
that, even at the bottom of the throat, the angular momentum quantum
numbers are not strictly conserved. In dual field theory language,
the attachment to a compact Calabi-Yau induces ceratin global
symmetry breaking perturbation. Ref.~\cite{DeWolfe:2004qx} estimated the leading
supersymmetry preserving operators to be of dimension 7.

This then
translates to the typical width of the lowest angular momentum mode
to be
\begin{equation}
m(m/M_{P})^6
\end{equation}
or equivalently to lifetime of such long-living relic
\begin{equation}
\left(\frac{M_P}{m}\right)^7\frac{1}{M_P} \label{giant}
\end{equation}
With an  inflation scale $m$ lower than $10^{12}$ GeV
 this could easily cause
a problem, since KK modes
live longer than 100 sec,
 their energy
release  destroys  BBN. If they live longer,  then there is
 too much dark matter content.

Furthermore, this problem is potentially much more severe
if the standard model is realized in another, longer throat with
such approximate isometries. As will be described in next section,
energy transfer from the inflation throat to a standard model
throat occurs via quantum processes of tunnelling and oscillation,
the other throat with standard model is likely to be of much
lower energy scale. The energy transfer then will involve
highly excited KK modes in the standard model throat and thus
will produce relatively large amount of the such angular KK modes.
With longer throat, the approximate isometry is more protected,
and lifetime of such angular KK Modes would be significantly longer.

We should warn the readers that, most likely, existence of such
approximate isometries is not generic. It is true that all known
Sasaki-Einstein manifolds which could play the role of $T_{1,1}$
of KS throats are equipped with some angular isometries, if smaller
than that of $T_{1,1}$. But this is probably result of limited
technology on our part in finding explicit examples. In dual
field theory language, such global symmetries are not required
supersymmetry in any intrinsic way. It is easy to envision
that generic throats in type IIB flux compactification involves
no such approximate isometries at all. Nevertheless our present
consideration should exclude certain subset of such inflationary scenarios
involving relatively symmetry throats.
In particular, the face value model based on KS throat
has problem of long-living angular KK modes.

\section{Issues with  Multi-Throat Scenarios}\label{sec:mult}

So far we studied in some detail how remnant of D-brane
annihilation cascade down to standard model sector, assuming that
a standard model is realized at the bottom of the inflation throat
as a brane world.
 Apart from potential problems with long-lived
relics associated with approximate isometries at the bottom of the
throat, realistic reheating seems possible, although its detail
differ from conventional reheating process. For one thing, the
reheating process involves two distinct phases, where localized
string modes and KK modes are first created and thermalized and
then later this energy is transferred to open string sector and
thermalize at  lower temperature.

SM throat requires the choice of warping factor $e^{-y_0} \sim 10^{-15}$.
On the other hand KKLMMT model suggest
inflationary throat to have
 $e^{-y_0} \sim 10^{-4 }$ to have right amplitude of
cosmological fluctuations.
This is one of the motivation
to consider the multiple throat scenarios.
This constrain can be relaxed if we admit
another source of cosmological inhomogeneities
related to the modulated cosmological fluctuations \cite{mod}.
Notice also that the scale of inflation is tightly
related to the SUSY breaking scale, and TeV scale gravitino
requires low scale inflation \cite{ar}.

This single-throat scenario is certainly the simplest possibility,
and some variant of it might work for real world provided that
supersymmetry generates further hierarchy from the inflation scale
down to TeV scale. On the other hand, with a single throat
scenario like this it is a little bit unclear how supersymmetry
would be broken at the right scale and how to generate the small
and positive cosmological constant which is observed in today's
universe. For model building purpose, it gives us more room to
consider flux compactification scenarios with more than one such
throats. Here we comment on new issues in reheating such a
multi-throat brane world.

One immediate fact is that there is never a thermal equilibrium
between any such pair of throats, due to the large redshift
factors. Classically the localized degrees of freedom in any of
the throat are effectively confined to the bottom of the throat,
and can interact with those in another throat only via highly
suppressed mixing operators. Wavefunction of such localized
particles of mass $m$ may have exponentially small tail toward top
of the throat $(m/M_{P})^d$ for some positive number $d$, so
interaction amplitude between two different throats should carry
suppression $(m/M_P)^{2d}$. In turn, the associated probability
goes like $(m/M_P)^{4d}$. Most crudely, the time scale of such
interaction is then
\begin{equation}\label{power}
\Delta t \sim \frac{1}{m}\left(\frac{M_P}{m}\right)^{4d}\label{unknown}
\end{equation}
A hint on what the number $d$ might be can be found in a work by
Dimpopoulos et.al. \cite{Dimopoulos:2001ui,Dimopoulos:2001qd},
which estimated tunnelling rate from one Randall-Sundrum universe
to another glued at the UV brane. The most optimistic estimate one
finds there is $d=1$.

We expect even lower tunnelling rate for
angular modes. This extra suppression for angular modes can also
be seen from the behavior of wavefunction outside the throats as
well. In between throats, that is, in  the ``big'' CY volume, we
can put the warping to be almost constant, $H(y) \sim H_0$, and
the wave equation reads as
\begin{equation}\label{eigen2}
\left[\partial^2_y+ m_{KK}^2R^2H_0^{1/2}-L^2\right]\Phi_{m_{KK};L}=0 \ ,
\end{equation}
Thus in between throats  solution of the KK wave equation (\ref{eigen})
is trivial, for $L=0$ modes it is simply a slow sinusoidal form
(in fact more or less a small constant since $m_{KK}R$ is very small),
while for angular KK modes it is exponential in $y$, $\Phi_{m_{KK};L}
\sim e^{\pm L y}$. As wavefunction traverses the classically
forbidden region, it will be more strongly suppressed
because of this behavior. For actual decay exponent with
smooth Calabi-Yau (rather than singular boundary condition),
we must do more careful analysis of the whole wavefunction.

While the rate of decay of angular KK modes is seemingly
thus more suppressed, this will  probably lead to an overestimate
of the lifetime of actual angular modes.  One problem is
that there is possibility that
higher angular modes will decay a pair of lower ones classically
without having to tunnel. To settle this,
one must study lowest lying KK spectrum
and $L^2$ dependence thereof. The other problem is that since the
angular isometries are approximate, the dominant decay channel
might be to change the angular momentum to lower value within the
same throat and then tunnel to another throat. For this, one must
take into account perturbation of type that lead to (\ref{giant}).
Thus, one should not take this estimation of the net decay rate
 too seriously for larger
values of $L^2$.

Let us return to the formula (\ref{power}).
With $d=1$, this
time scale is clearly much larger that $1/H\sim M_P/m^2$ at the
exit of inflation, and thus the two throats are not in thermal
equilibrium. While the $1/H$ increases with later evolution, so
does $\Delta t$ at much faster rate. Any exchange of energy
between two such throats are possible only via possible decay of
heavy particles in one throat into lighter ones in the other.

This warns against one easy mistake in dealing with multi-throat
cases. In early universe one often invokes equipartition principle
and assume that energy is deposited in each and every degree of
freedom in equal amount.  A variant of this may be employed to
argue that, if throat 1 comes with many more degrees of freedom
than throat 2, then more energy is deposited to throat 1. However
we are also familiar with the fact that some subsystem can be
frozen away from thermal equilibrium and will evolve on its own if
its interaction rate with the rest is much   smaller  that $H$.
The above estimate of interaction between distinct throats tells
us that such naive counting based on intuition from thermal
equilibrium physics cannot be trusted.

One can make this a little more precise using a simple quantum
mechanics. Consider throat 0 which drove inflation and contains
massive KK modes of scale $m$ after branes annihilation, and a
longer throat 1 with mass scale $\mu\ll m$. The dominant decay
channel of heavy particle of mass $m$ in throat 0 is found by
realizing that, since the two throats are not completely
disconnected, KK modes of throat 0 will mix and oscillate quantum
mechanically with KK modes of throat 1 with similar mass, say
$m'\sim m \gg \mu$. With a generic form of the mixing mass matrix,
\begin{equation}
H=\left(\begin{array}{cc}m & \epsilon \\ \epsilon &
m'\end{array}\right)
\end{equation}
we expect $\epsilon\sim m(m/M_P)^{2d} \ll m$. The oscillation of
the initial state in throat 0 will induce amplitude of KK modes in
throat 1 which is roughly
\begin{equation}
\langle 1|e^{-iHt}|0\rangle\sim e^{-imt}\,\frac{\epsilon}{\Delta
m}\;\sin(\Delta m t)
\end{equation}
with $\Delta m\simeq m_1-m_2$. Part of the quantum state
oscillating into the throat 1 will then decay since there are many
lighter states in throat 1. Naively one may think that having many
decay channel in throat 1 is very helpful in transferring energy
into throat 1 from throat 0, since there are a lot of phase volume
in throat 1.

\begin{figure}
\begin{center}\leavevmode\epsfxsize=0.5\columnwidth \epsfbox{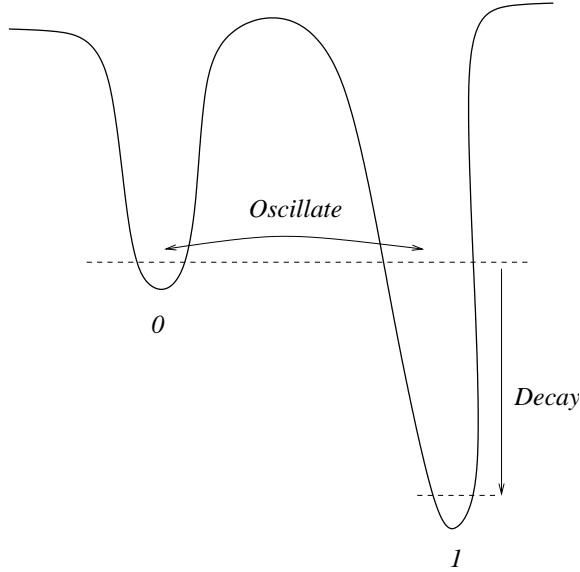}\end{center}
\caption{KK modes in the inflation throat can decay to another throat via
quantum oscillation. While the oscillation amplitude depends on mass
eigenvalue distribution in the 2nd throat, it will be also suppressed
further if the state in throat 1 has a large width.}
\label{fig:mix}
\end{figure}

However, to maintain consistency, the above mass matrix must be
modified to reflect  the decay width of the state in throat 1.
\begin{equation}
\tilde H=\left(\begin{array}{cc}m & \epsilon \\ \epsilon &
m'-i\Gamma\end{array}\right)
\end{equation}
With $N$ number of states available as decay products in throat 1,
the width should scale like $\Gamma \sim \lambda m' N$ with a
small but finite coupling constant $\lambda$. With the new $\tilde
H$ that has the decay process built in, we can estimate how fast
the initial state in throat 0 lose its amplitude,
\begin{equation}
e^{-i\tilde H t}|0\rangle\sim e^{-\tilde \Gamma
t}e^{-imt}\left((1-O(\epsilon^2))|0\rangle+
\frac{\epsilon}{\Gamma-i\Delta m}\sin{\Delta m t}|1\rangle\right)
\end{equation}
where the effective width
\begin{equation}
\tilde \Gamma\equiv \frac{\epsilon^2}{\Delta m^2+\Gamma^2}\Gamma
\end{equation}
of the state $|0\rangle$ due to the
oscillation
is actually suppressed as the width $\Gamma$ of state $|1\rangle$
increases.

While somewhat counter-intuitive,
this can be understood from the basics of the quantum
oscillation. Two states mixes with each other well if they are
almost degenerate while the mixing is suppressed by the mass
difference. What this simple computation tells us is that the
suppression depends on difference in the complex mass.
Another way to see this is that the decay of state $|0\rangle$ occurs
at second order in the perturbation theory.
It must first oscillate into state
$|1\rangle$ and experience the decay width $\Gamma$ and then come
back to $|0\rangle$. The standard 2nd order perturbation due to
the mixing $H_{01}=\epsilon$ gives
\begin{equation}
{\rm Im}\frac{\langle 0|\tilde H_{01}|1\rangle \langle 1|\tilde
H_{10}|0\rangle}{\tilde H_{11}- \tilde H_{00}}
\end{equation}
as the width, which is exactly  $\tilde \Gamma$.

Finally we must
take into account that there are also roughly $N$ number of states
that can be used for such oscillation, so the net width of state
$|0\rangle$ is more like
\begin{equation}
N\tilde \Gamma
\end{equation}
Assuming large $N$, so that $\Delta m$ is smaller than $\Gamma$,
we find
\begin{equation}
N\tilde \Gamma\sim \frac{\epsilon^2}{\Gamma}N\sim
\frac{\epsilon^2}{\lambda m}
\end{equation}
which shows that, to leading approximation, the effect of having a
large number of state in the low energy throat completely washes
out.\footnote{This estimate would not apply to relatively stable
state of lower mass $\sim \mu$ in throat 1 at its bottom. However the
corresponding $\epsilon$ would be far more suppressed at $\sim
(m\mu/M_P^2)^d$, since the state $|1\rangle$ in that case would be
living at the bottom of throat 1.}
Any throat that is significantly longer that the inflation throat
would receive energy as dictated by the mixing of typical KK
wavefunctions, irrespective of how many decay channel each comes
with. In particular, having a large number of open string decay
channel in throat 1 will make matters worse, since these will
contribute to $\Gamma$ but not to the mixing.

\begin{figure}
\begin{center}\leavevmode\epsfxsize=0.5\columnwidth \epsfbox{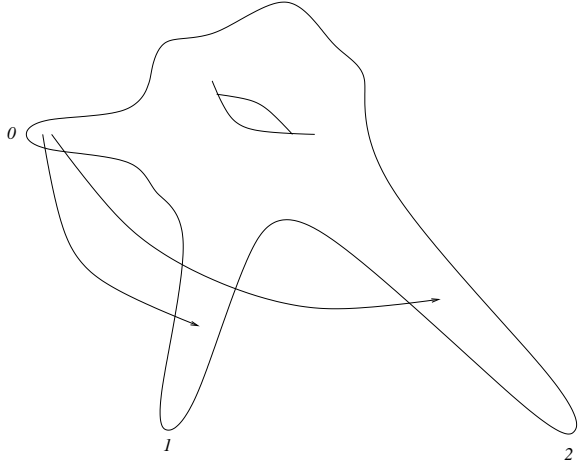}\end{center}
\caption{KK modes in the inflation throat deposit energy to lower
energy throats. The branching ratio will be largely determined by
how throats are distributed in the internal manifold with respect to
the inflation throat and less sensitive to the field content of
each throat. Energy deposited in throat 1 will later decay to
throat 2, but at much more suppressed rate because its mass
scale is far lower than that of the inflation throat.} \label{fig:tunnel}
\end{figure}

Such a universal nature of decay into different throats will make
the reheating process quite delicate. For instance suppose the
compact manifold involves three different throats. We will label
them as 0 the inflation throat, 1 the SUSY breaking throat, and 2
the standard model throat, with the mass scales $m\gg \mu_1\gg
\mu_2$, possibly with large number of additional open string
degree of freedom in throat 2. Above consideration tells us that
energy deposit into throat 1 and throat 2 are largely determined
by each mixing mass matrix element $\epsilon$'s and initial energy
level $m$, and independent of $\mu_{1,2}$. This would result in
significant deposit of energy in throat 1 as well as in throat 2,
provided that the process is fast enough. In terms of (\ref{unknown}),
$d=1$ with conventional inflation scale would suffice.

With $\mu_1$ at some intermediate scale,
furthermore, whatever energy deposited there will behave as a
massive particles as universe expands and cools down. Compared to
the radiations in throat 2 (mostly photons and other light
standard model degrees of freedom), the density of such massive
particles dilute rather slowly.
 Even a small amount of deposit in
throat 2 will quickly overtake and dominate energy density of
universe rendering reasonable BBN  impossible.
While energy deposited in throat 1 will eventually decay to throat 2,
its decay width would be dictated by $(\mu_1/M_P)^{4d}$ which
is much smaller that $(m/M_P)^{4d}$. Relics in throat 1 would
be very difficult to remove. We
believe that this would cause a very serious problem whenever we
build models based on KKLMMT-like inflation with warp factors
playing crucial roles in hierarchy, determination of cosmological
constant, and low scale of supersymmetry breaking.

Notice that this problem with low-laying KK modes is
independent from the angular KK modes. While problem with
angular KK modes is there only if the base manifold of
the throat happens to be quite symmetric, this problem
is much more generic, provided that there are some
intermediate-length throat containing a hidden sector.

\section{Summary}\label{sec:summary}

It is fair to say that
string theory inflation is at a crossroads.
There are several potential directions for building realistic
inflationary models. The end point of all inflationary models is reheating
of the universe.
Successful reheating means almost complete conversion of inflaton energy into
thermalized radiation without any dangerous relics, in order to
provide a thermal history of the universe compatible with
 Big Bang Nucleosynthesis (BBN), baryo/leptogenesis,  and other observations.
All of this provides us with constraints on string theory  inflationary scenarios.

In this paper we investigated reheating after brane-anti brane annihilation.
The starting point was the KKLMMT model of warped brane inflation
with a background KS geometry.
Control of the calculations is possible in the regime of
low curvature KS geometry with all scales (like the
radius at the bottom of the throat) larger than the string scale.
Apart from the justification of
computation, we do not have much reason to believe these constraints. In fact,
for large $R_+/R_-$ to generate hierarchy, one may prefer $R_-$
nearer to the string scale \cite{CMP}.
KS geometry also admits isometries of the inner manifold.
Another important parameter is the warping factor which defines the
energy scale of inflation. The single throat scenario with low energy inflation
just above the SM scale is the simplest  possibility if there are alternative mechanisms
to generate
 cosmological fluctuations (like modulated fluctuations).
Otherwise we can have a multiple throats scenario with inflation and
the SM sector
located in different energy scale throats.
Thus, the models bifurcate into single throat or multiple throat scenarios,
with different ranges of parameters to stay in or out of low curvature
  geometry.

We followed the way the energy of annihilated branes transfers into SM radiation.
The first step after annihilation is the excitation of closed string loops,
located near the bottom of the throat, where the local string mass scale is
reduced by the warp factor  $e^{-y_0}$.
They quickly decay into 10d gravitons (and particles in their supermultiplets).
 From a four dimensional perspective, 10d gravitons are manifested as
usual 4d gravitational radiation and
massive KK modes.

The generation of 4d gravitational radiation is a universal prediction of
string theory brane inflation. In principle, it is a source of concern
because BBN excludes more than few percent of radiation in forms
other than photons and 3 light neutrino species.
 Massless radiation relics today are photons with the fraction of the total
energy density $\Omega_{\gamma} \simeq 5 \times 10^{-5}$
 and very light neutrinos with $\Omega_{\nu} \simeq 1.6 \times 10^{-2}$.
Successful BBN potentially allows an extra few percent of  energy density
in light species, for instance background relic gravitons.  Hence the energy density
of relic gravitons cannot exceed the limit
 $\Omega_{GW} \leq   5 \times 10^{-6}$.
For instance, in brane inflation models without warping
and just a pair of $D3-\bar D3$ branes, overproduction of
gravitons is a problem, which can be cured by annihilation of $\bar D3$
with a stack of $N$ $D3$ branes to dilute the abundance of gravitons by $1/N$.
In warped geometry, the situation with gravitons is very different.
The decay rate of closed loops  into gravitons (i.e. homogeneous KK modes)
is suppressed by $M_p$, while decay into KK modes is suppressed by
the local string mass $e^{-y_0}/\sqrt{\alpha'}$ at the bottom of
the throat, which is much larger than $M_P$.
As a result, the fraction of energy
deposited into gravitons is reduced by the factor $e^{-2y_0}$.
For inflation at $10^{14}$ GeV  this is $e^{-2y_0} \sim 10^{-8}$.
We therefore conclude  that\\
 {\it the warped brane inflation scenario
predicts a residual background  of relic gravitational radiation
at the level  $\Omega_{GW} \sim e^{-2y_0}$. The wavelengths are located
around the energy scale of closed loop excitations after inflation,
redshifted by the expansion of the universe, and up to some numerical factors
close to that of the relic photons, around $mm$ scale.}

The bulk of the energy after decay of closed string loops is in KK modes.
They interact and reach thermal equilibrium with a temperature
comparable with the inflation energy scale.
A much more serious concern comes from the angular KK modes.
Indeed, the 6d compact manifold in the KS throat geometry has isometry,
near the bottom $S^3 \times S^2$ isometry.
Conservation of angular momentum associated with these isometries
forbids complete decay of angular KK modes, but allows only their
annihilation. As a result a residual amount of massive angular KK modes
freezes out in an expanding universe, posing a problem for the scenario.

In the context of CY compactification isometries are only approximate,
which allows the angular KK modes to decay.
However, the width of their decay is suppressed by the factor $(m/M_p)^7$.
While this gives a fast enough decay time for angular KK modes in
$10^{14}$ GeV inflation, it is too long a time for angular KK modes in
the SM throat.

In fact,
there are two independent KK mode problems. One has
to do with the approximately conserved angular momentum which we just discussed.
 This is there
for both the single and multiple throat cases as long as a long throat has
an approximate isometry.
The other problem comes  from a low-lying KK mode created in some
intermediate throat (low-scale inflation throat
or susy-breaking hidden-sector throat)
in multi-throat scenarios. Their
decay is via tunnelling only, and is thus suppressed.
This suppression has little to do with any isometry and is thus more
problematic, at least for generic  multi-throat cases.

It is an interesting exercise to calculate the cosmological abundance of residual KK modes,
which we considered only briefly for the single throat case in Section 4.
Let us mention the cosmological relevance of the issue.
The energy density of massive relics $\Omega_{KK}$ dilutes slower than that
of radiation, therefore  $\Omega_{KK}$ sooner or later begins
 to dominate the expansion of the universe. Special conditions are needed to
tune   $\Omega_{KK}$ to be the Cold Dark Matter component, to constitute today
 $\Omega_{CDM} \simeq 0.27 $. If KK  particles are abundant
in the very early universe
it  becomes matter dominated too early to allow successful BBN.
This is a typical situation with SUSY moduli fields and gravitinos.
A similar problem is inherited by simple versions of
rolling tachyon inflation \cite{KL}.
On the other hand, it is fair to keep a possibility that
$\Omega_{KK}$  is a candidate for the CDM, if KK particles are
long-lived and are created in exactly the right
amount \cite{KK}.

\vspace{1cm}

{\bf Acknowledgements}. We thank J. Cline, G.Felder, K. Hori,  S. Kachru, R. Kallosh,
 K. Lee, A. Linde, R. Myers,  A. Peet, G. Shiu,  O. Saremi, and S. Trivedi for useful
discussions. PY is grateful to the Fields Institute, to Perimeter Institute,
and also to the Canadian Institute  for Theoretical Astrophysics, for kind
hospitality. LK was supported by CIAR
and NSERC. PY is supported in part by Korea Research Foundation
Grant KRF-2002-070-C00022.

\end{document}